\documentclass[conference]{IEEEtran}
\IEEEoverridecommandlockouts
\usepackage{cite}
\usepackage{amsmath,amssymb,amsfonts}
\usepackage{algorithmic}
\usepackage{graphicx}
\usepackage{textcomp}
\usepackage{xcolor}
\usepackage{booktabs}
\def\BibTeX{{\rm B\kern-.05em{\sc i\kern-.025em b}\kern-.08em
    T\kern-.1667em\lower.7ex\hbox{E}\kern-.125emX}}

\title{Whisper-Based Speech Transcription from Videos Across Multiple Languages for Cross-Cultural Understanding \\
\thanks{This work was partially supported with funding from the Defense Advanced Research
Projects Agency (DARPA) Cross-Cultural Understanding (CCU) program under Contract No. HR001122C0034 
and from the Institute of Information \& Communications Technology Planning \& Evaluation (IITP) 
with a grant funded by the  
Ministry of Science and ICT (MSIT) of the Republic of Korea in connection with the Global AI Frontier Lab International Collaborative Research Agreement. (Nos. RS-2024-00469482 \& RS-2024-00509279).}}
\author{\IEEEauthorblockN{Michael Picheny}
\IEEEauthorblockA{\textit{NYU Courant Institute  School of Mathematics, Computing, and Data Science} \\
\textit{NYU}\\
New York, USA \\
map22@nyu.edu}}

\begin{document}
%\ninept
%
\maketitle
\begin{abstract}
Cross-cultural understanding has become increasingly important in today's highly connected, cross-national world. The success of LLM-based technologies is now driving the development of automated tools to aid understanding for non-native people trying to succeed in cross-cultural environments. Building such automated tools is often done by leveraging in-the-wild text, audio, and video data. This paper presents techniques for improving speech recognition-based transcript creation in multiple languages from videos to better train these automated tools. The focus is on processes and speech tools that can easily be used by cross-cultural tool builders without requiring deep speech processing expertise. Using publicly available videos from YouTube and Whisper-based tools, average transcription error rate across seven languages (Spanish, Japanese, Korean, Mandarin, Turkish, Russian, and Hebrew) of ~30\% are observed. With a modest amount of fine-tuning data, the average error rate can be reduced to ~20\% making such output much more usable for downstream processing. Speech and metadata associated with these videos that can be used by the community to further refine these experiments are released as well.     
\end{abstract}
\section{Introduction}
\label{sec:intro}
Cross-cultural understanding has become increasingly important in today's highly connected, cross-national world. The success of LLM-based technologies is now driving the development of automated tools to aid understanding for non-native people trying to succeed in cross-cultural environments. Building such automated tools is often done by leveraging analyses of multimodal data, including text, audio, and video data \cite{li2023normdial, fung2023normsage, wang2024normgenesis, yuan2024measuring, sahu2025minds}.

One way to leverage audio data for such purposes is to convert the audio data into text data and then utilize text-based LLM-based tools to extract relevant cultural information (see \cite{pawar2025survey} for a survey of such tools). The degree to which audio data can be leveraged depends on the performance of the extraction process. Deep learning has greatly improved speech recognition performance over the past several years. Open-source leaderboard results \cite{srivastav2025open} might suggest to the casual user of speech recognition that word error rates are now significantly below 10\% across a variety of tasks, and therefore not impact research on cultural understanding.  However, these tasks tend to  contain prompted and/or stylized speech from professional speakers and are recorded in relatively benign environments. For cultural understanding, the focus is on information extraction from naturally produced speech "in-the-wild". Extracting accurate transcripts from such realistic audio data can still be challenging, especially for languages that are not as data-rich as English. Accuracy can be improved by building custom systems leveraging thousands of hours of data, but such approaches are not often feasible for researchers interested in cross-cultural understanding. Such researchers do not tend to be speech experts and would prefer easy, inexpensive methodologies for extracting transcripts from audio. Finally, because cultural markers are low-density with respect to hours of audio, thousands of hours of audio need to be analyzed to produce an adequate amount of data to create downstream cultural understanding models, making processing time and cost an important practical issue.

Given the increasing interest in research in developing automatic tools for cross-cultural understanding, and that an important pre-processing step is the extraction of audio transcriptions from large amounts of multimodal, multilingual data, this study was initiated with the following three goals. The first was to estimate open source speech transcription accuracy for data relevant to cross-cultural understanding across a variety of languages. The second was to outline a process that non-speech researchers with limited resources could realistically use to collect data and build usable speech transcription systems for cultural understanding. The third goal was to share representative speech data with the community to spur new research to obtain further performance improvements.   

Given ease-of-use, processing time,  and cost constraints Whisper-based \cite{radford2023robust} speech transcription generation was utilized. A flavor of Whisper developed at the Univerity of Oxford called WhisperX \cite{bain23_interspeech} seemed particularly well suited for this purpose. Whisper itself is a high-performing open source speech recognition system with good multilingual performance and fine tuning capabilities. It has an easy to use API. WhisperX incorporates a number of significant speedups to basic Whisper processing and also is capable of speaker diarization (important for conversation analysis). It also has an easy to use API. 

The rest of this paper describes the detailed methodology developed that others could potentially copy and use to extract transcripts in large quantities from long-form speech. 
Section \ref{sec:choice} describes the choice in recognizers, Section \ref{sec:lmdata}, the choice of languages, Section \ref{sec:pipeline}, the processing pipeline, Section \ref{sec:selection}, the selection of videos, Section \ref{sec:experiments}, out-of-the-box and fine tuning experiments, Section \ref{sec:discussion}, discussion of results including usability implications, and Section \ref{release}, an overview of the data and metadata release.

\section{Data Processing}
\label{sec:data}

\subsection{Recognizer Choice}
\label{sec:choice}
Whisper-based processing was chosen because of an easy to use API and excellent out-of-box performance. Whisper was trained on 680,000 hours of audio. 563,000 hours are English, and 117,000 hours cover 96 other languages \cite{radford2023robust}.  At one point in time, 2000 hours of speech was viewed as an enormous amount of training data. 13 of the 96 languages contain at least 2000 hours of audio, so good out-of-box Whisper performance is not unexpected for these languages. Unfortunately, no details are given about the sources of the data.

Both Whisper and a derivative ("WhisperX") were evaluated. WhisperX was developed at the University of Oxford. It not only performs Speech Recognition but also performs speaker diarization by incorporating an open-source modular package for diarization (Pyannote \cite{Plaquet23, Bredin23}). Pyannote comes with a set of pretrained models for English but can be trained from scratch for other languages as well. For research on cultural understanding, speaker diarization is a key feature as it is needed for identification of speaker turns in a conversation. WhisperX also contains a sped-up version of Whisper. The speedup is essential for processing the thousands of videos needed to extract enough information to train downstream models for cultural understanding. All ou experiments utilized the "large-v2" Whisper model (\cite{bain23_interspeech} claims WhisperX is 10x faster than Whisper for the large-v2 model). For some languages, the "large-v3" models might have had improved performance, but Github discussions (e.g., \cite{whisper_github_discussion_2024}) suggested that large-v3 might be more prone to hallucinations for noisy data. 

\subsection{Language Data}
\label{sec:lmdata}
The choice of languages to study was inspired by the languages selected for the DARPA Computational Cultural Understanding program (CCU) \cite{DARPACCU} (Mandarin, Spanish, Korean, Japanese, Russian, and Turkish). In CCU, speech recognition was needed to extract transcripts from audio and video data. Since speech recognition in itself was not a CCU focus no manual transcripts were supplied. It was therefore not possible to evaluate speech recognition performance. 

The 6 CCU languages were each represented by over 4000 hours of speech (Table \ref{tab:whisper}) in Whisper training data, so good out-of-box performance seemed to be a reasonable expectation. Hebrew was added as a challenge language because its coverage in Whisper relative to the other six languages is significantly less (688 hours). This is large enough to expect decent performance, but its behavior relative to the other languages before and after fine-tuning might be different. In addition, because it was familiar to the author, it made it easier to debug the various tools and pipelines needed to process the data. 

\begin{table}[tb]
    \centering
    \caption{Size of Whisper training data for the selected languages.}
    \centering
    \label{tab:whisper}
   
    \begin{tabular}{lr}
        \toprule
        \textbf{Language} & \textbf{Hours} \\
        \midrule
        Mandarin & 23446 \\
        Spanish & 11000 \\
        Russian & 9761 \\
        Japanese & 7064 \\ 
        Korean & 7793 \\
        Turkish & 4333 \\
        Hebrew & 688 \\
        \bottomrule  \\
\end{tabular}
\end{table}

\subsection{Processing Pipeline}
\label{sec:pipeline}

\begin{table*}[t]
    \centering
    \caption{Consolidated Language Metadata (Duration in Hours)}
    \label{tab:langstats}

    \begin{tabular}{lllccccccc}
    \toprule
    \textbf{Size} & \textbf{Type} & \textbf{Stat} & \textbf{Man.} & \textbf{Spa.} & \textbf{Rus.} & \textbf{Jap.} & \textbf{Kor.} & \textbf{Tur.} & \textbf{Heb.} \\ \midrule
    
    %% --- Size: All ---
    All & Train & Num Videos & 127 & 20 & 186 & 82 & 247 & 740 & 135 \\
    All & Train & Num Segs & 57819 & 9063 & 72979 & 10465 & 48319 & 205018 & 61210 \\
    All & Train & Duration & 50.35 & 8.45 & 117.30 & 31.01 & 53.58 & 207.22 & 63.88 \\
    \cmidrule{1-10}
    All & Dev & Num Videos & 10 & 3 & 6 & 8 & 13 & 29 & 8 \\
    All & Dev & Num Segs & 3738 & 381 & 1142 & 1007 & 1679 & 8207 & 5266 \\
    All & Dev & Duration & 3.37 & 0.31 & 3.57 & 2.18 & 1.68 & 8.36 & 5.18 \\
    \cmidrule{1-10}
    All & Test & Num Videos & 21 & 10 & 12 & 11 & 23 & 58 & 13 \\
    All & Test & Num Segs & 7572 & 1713 & 7464 & 1354 & 3391 & 15317 & 6758 \\
    All & Test & Duration & 7.65 & 1.58 & 8.23 & 5.72 & 4.21 & 15.10 & 7.44 \\

    \midrule

    %% --- Size: Small ---
    Small & Train & Num Segs & 5782 & 9063 & 6999 & 3488 & 8053 & 10250 & 6121 \\
    Small & Train & Duration & 5.01 & 8.45 & 11.04 & 10.35 & 8.98 & 10.34 & 6.32 \\
    \cmidrule{1-10}
    Small & Dev & Num Segs & 373 & 381 & 1142 & 336 & 560 & 820 & 526 \\
    Small & Dev & Duration & 0.32 & 0.31 & 3.57 & 0.74 & 0.57 & 0.83 & 0.52 \\
    \cmidrule{1-10}
    Small & Test & Num Segs & 757 & 1713 & 746 & 451 & 1130 & 1531 & 675 \\
    Small & Test & Duration & 0.75 & 1.58 & 0.83 & 1.88 & 1.40 & 1.51 & 0.74 \\

    \bottomrule
    \end{tabular}
    
\end{table*}

A significant fraction of the CCU data came from YouTube. This suggested the following data collection process. The YouTube platform processes HTTP GET requests where state or filter criteria are passed via query parameters \cite{dataapi} and search parameters \cite{serpapi} within the Uniform Resource Identifier (URI). One can use these parameters to specify search keywords, and also locate videos that match particular criteria, such as restricting videos to the current year, month, week, or day, videos with subtitles, and videos provided under a Creative Commons license. Third parties have developed easy-to-use command-line interfaces to access YouTube data. yt-dlp \cite{yt-dlp} leverages the above mechanism to search for videos, return content, and help interpret the returned content in a user friendly fashion. The Jtubespeech repository \cite{takamichi2021jtubespeechcorpusjapanesespeech} is a set of tools built on top of yt-dlp that make it easy to collect large amounts of Youtube-based speech data for specific languages. 

The supplied CCU videos from Youtube (provided via YouTube video ids) were filtered using the Jtubespeech and yt-dlp tools to select those videos with manually created subtitles and provided with a Creative Commons license. Three of the languages (Mandarin, Korean and Turkish) had little or no videos that satisfied these criteria. For these videos, and for Hebrew, processing was started from scratch. The time period for crawling for new videos was restricted to the previous year (mid 2024- mid 2025) except for Mandarin, where too many videos were being produced, so the time period was restricted to April, 2025 (Note the "large-v2" Whisper model was released in early 2023). 

Here is an outline of the processing steps. Details are provided as part of Goal 2 - to provide a process for non-speech experts to build transcription systems. All the processing was performed using the Jtubespeech github repository tools. The repository documentation is good and the tools easy to use. For those languages that are CCU based the first two steps were unnecessary. The associated Jtubespech tool for each step is provided in boldface.

\begin{enumerate}
\item Extract a list of words created from the titles stored in a Wikimedia index dump file, a file with the titles of recent Wikipedia articles for different languages (index dump format described in \cite{Wikimedia}) ({\bf make\_search\_word}).
\item Use this list of words to search for videos that contain subtitles and are creative commons licensed and are restricted to a specific time period  ({\bf obtain\_search\_word})
\item Extract information about whether or not the subtitles for a video are automatic or manually produced. ({\bf retrieve\_subtitle\_exists})
\item Download the videos with manual subtitles along with the subtitles.  Extract the audio and change the sampling rate to be 16 KHz mono-channel audio. ({\bf download\_video})
\end{enumerate}

The Jtubespeech tools were modified so that they could process data in multiple parallel batches from one master list of video ids. The tool in step 1 was modified to limit titles to those that only contained words for the language in question (the titles often contained English). Note that this required language dependent processing, typically restricting the numeric values of characters in a UTF-8 representation to lie in specific ranges appropriate for a specific language. The original search parameters (found in the {\bf util.py} tool in the {\bf make\_query\_url} function) were modified to restrict the time and the license type, see \cite{serpapi} for the values of the search parameters and the options.  

The above process produced captions for each video as a vtt file \cite{vttfile}, a format suitable for closed-captioning the videos. A vtt file contains the text, the beginning, and the end time of the caption to be displayed. They tend to be just a few seconds long and abut each other closely in time. The time marks in short segments were frequently inaccurate for the purpose of marking audio word and/or phrase/sentence boundaries. To ameliorate this problem, the following procedure was used.

\begin{enumerate}
\item Merge consecutive segments from the vtt file till either 20 seconds of audio is accumulated or a measurable silence gap (.1 sec was chosen) between segments is found. In vtt files, adjacent segments abut in time unless there was some significant silence between the segments. This process produced either long segments or silence-delimited segments, both of which would allow for better alignments to be computed.   
\item For each new merged segment, use one of the supplied language dependent phonetic alignment models provided by WhisperX to determine more accurate begin and end times for the words in each merged segment. 
\item Resegment each merged segment  using the modified word times from step 2 at silences $>=$ .5 sec  producing a new set of segments.
\end{enumerate}

While far from perfect, this process appeared to produce better time alignments than those provided in the original vtt files. 

\subsection{Video Selection Process}
\label{sec:selection}

The new segments produced via the above process were then transcribed using WhisperX and scored using the NIST scoring toolkit SCTK \cite{SCTK} for computing word and character error rates. The new segment transcripts were used as references.  This produces an overall Word Error Rate for each video transcript. Videos with a WER $>$ 50\% were discarded; too high a WER threshold would result in data whose transcripts were likely to be of poor quality. Videos shorter than 5 minutes and longer than two hours were also discarded. The former hopefully would improve the chance of selecting videos with the potential of containing interesting cultural markers. The latter was a practical constraint, as longer videos could not be easily processed without significant code changes. The set of associated audio files were divided into separate  training, development, and test files, with the bulk of the data for each language assigned for training. The fractional split of the videos across each language varied because the total number of available video files varied, with some languages having relatively few files. The average duration per file across languages varied as well. A "small" version of the data was also created with about 10\% of the data to streamline fine-tuning experiments given limited computational resources. 

The breakdown across languages  are shown in Table~\ref{tab:langstats}. There is significant variability across the languages in the amount of data that was obtained. When possible the original CCU data was used when there were  enough files tagged as having Creative Commons licenses (10 hours minimum) (Spanish, Russian, Japanese), otherwise (for Mandarin, Korean, Turkish, Hebrew) Youtube was crawled for more data as described above. 

A significant assumption is that the new crawled videos are as suitable for cultural understanding purposes as the original curated CCU data. By limiting the video durations to be greater than 5 minutes the hope was to obtain videos containing interactions and some amount of actual cultural markers. Note that CCU involved processing literally thousands of videos per language to ensure enough cultural markers were obtained to train good models and allow for in-depth evaluations, as cultural markers tend to be low density with rspect to the amount of audio transcribed.   

\section{Experiments}
\label{sec:experiments}
Three sets of experiments were run on all seven languages. The first set of experiments tested recognition performance with Whisper and WhisperX "out of the box". The second set of experiments fine-tuned Whisper and WhisperX using the "small" set of training data (Table \ref{sec:selection}). The third set of experiments was similar to the second except that fine-tuning was performed on the "all" (complete) set of training data (Table \ref{sec:selection}). For the purpose of rapid experiment turnover the "small" set of test data was used for recognition, and for fine tuning, the "small" set of development data was used across languages. Word error rates for all languages were computed except for Mandarin and Japanese, where character error rates were used. All fine tuning and inference was performed using Pytorch 2.7.1 and the torch-based Seq2Seq tools using whatever gpus were available on the available high performance compute cluster, (usually A100s, sometimes RTX6000s). 

\subsection{Out-of-Box performance}
\label{sec:oob}
\begin{figure}[tb]
\centering
\includegraphics[width=\columnwidth]{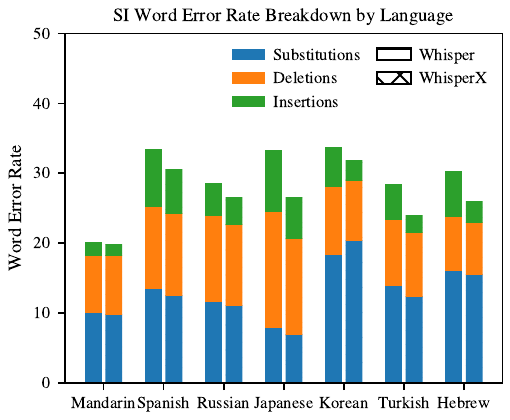}
\caption{Out-of-the box performance for Whisper vs WhisperX}
\label{fig:oob}
\end{figure}

Figure~\ref{fig:oob} presents out of box performance across languages for Whisper and WhisperX. WERs tend to lie between 25\%-30\%. There is a slight edge in performance for WhisperX; it is also considerably faster than Whisper itself. Mandarin error rate is substantially lower than the other languages,  probably because of the use of character error rate (CER) as a metric. However, Japanese error rates, also computed using CER, seem high, with the bulk of the errors arising from a high deletion error rate.  No obvious difference is  seen in overall WER for CCU vs non-CCU data, though the balance of errors across substitutions, deletions, and insertions across the two data sources are different. Similarly, no obvious difference is seen for Hebrew, with less Whisper training data, and the other languages.    

\subsection{Performance after Fine-Tuning}
\label{sec:ft}
The next set of experiments focused on fine tuning. Initial experiments suggested that simply unfreezing all the parameters in the model and running two epochs of training on top of the base models with a learning rate of 1e-5 and weight decay of .005 was as good, if not better, than anything else tried.  However, no attempt was made to perform an in-depth search to find the optimal (single) processing recipe across languages. 

\begin{figure}[tbh]
\centering
\includegraphics[width=\columnwidth]{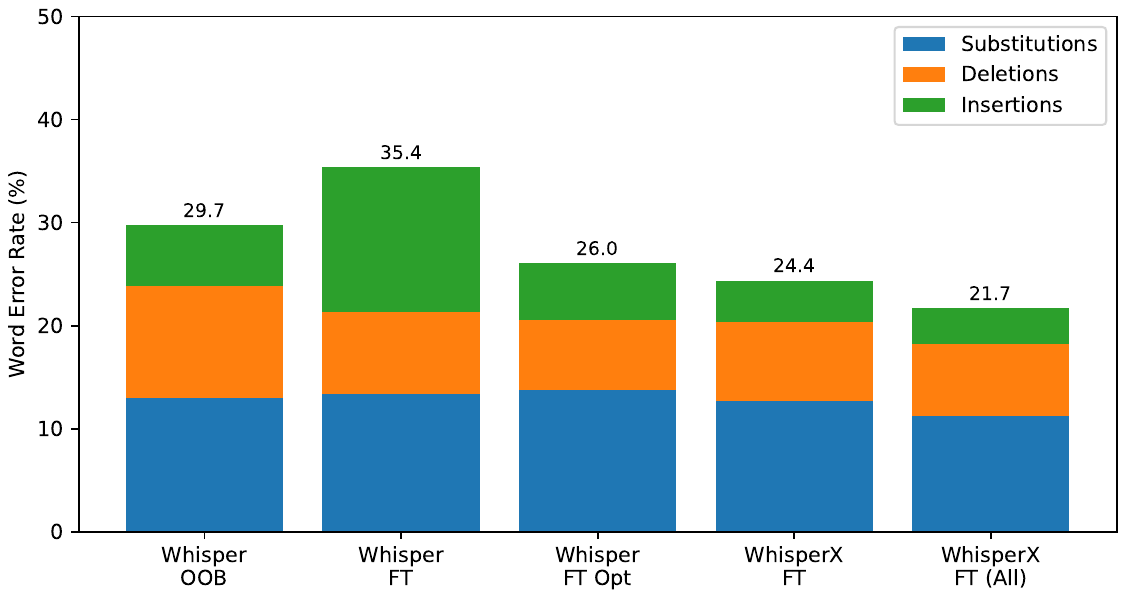}
\caption{Performance after Fine-Tuning}
\label{fig:ft}
\end{figure}

Figure~\ref{fig:ft} presents out of box performance averaged across language. The first bar is Whisper performance out of the box (29.7\% WER). The second bar is the result of fine-tuning using the "small" amount of data setting found in Table~\ref{tab:langstats} for each language. Performance after fine-tuning is {\bf worse} than out-of-box performance! Perhaps this occurred because of poor parameter tuning, or not using more sophisticated adaptation schemes such as LORA \cite{song24_interspeech}.  However, it can be seen that the main reason for the WER increase was a significant increase in the number of insertions. The insertion increase was almost always due to an increase in hallucinations. One simple technique used to reduce hallucinations is to reduce the maximum number of tokens that can be generated by explicitly specifying {\it generation\_config.max\_new\_tokens},  one of the arguments to the Seq2Seq trainer code. A setting of 64 resulted in output transcript truncation. A setting  of 256 seemed to generate the best results given the maximum length of a test segment was limited to 20 seconds. The third bar presents the result of this setting for Whisper based decoding (26.0\% WER). As can be seen, some improvement with fine tuning now results relative to out-of-box performance (29.7\%).

For WhisperX, the corresponding WER number to the Whisper results (26.0\%) is the fourth bar (24.4\%). No attempt was made to tune any of the WhisperX parameters. Note WhisperX was less prone to hallucinations. This is perhaps due to a much better VAD used in WhisperX relative to Whisper, resulting in elimination of more low-level noise, and shorter audio segments to recognize, both of which tend to reduce the amount of hallucination. 

The third set of experiments fine-tune on all the training data. Only results for WhisperX are presented (Whisper results were slightly worse). The final result is shown in the fifth bar, achieving another decrease in average WER (21.7\%).

Statistical significance was checked for the above results by applying paired bootstrap tests \cite{bisani2004bootstrap} to all 10 pairwise condition comparisons, aggregated across all 7 languages. All comparisons were significant (p $<$ .001) except for Whisper out-of-box performance vs. Whisper optimized fine tuning performance (p $=$ .012). This suggests that the "small" amount of data used for fine tuning - 10 hours or less per language - was just too small to generate large enough improvements to be considered significant at the desired threshold.  

No substantial difference in the performance of Hebrew relative to the other languages before and after fine tuning was observed even though it was represented by much less training data in Whisper than the other six languages. It was also noted that performance for the four languages harvested from YouTube was substantially better than the three languages whose was usable (in terms of having Creative Commons licenses) from the CCU program. This is perhaps attributable to the careful screening process used by LDC in data selection to ensure the data was appropriate for cultural understanding purposes, potentially resulting in more complex data. 

\section{Discussion}
\label{sec:discussion}
The above results demonstrate that the average word error rate for transcription of difficult speech using state-of-the-art open source speech tools and systems and relatively light (compared to industrial strength)  computing resources can be lowered by almost 33\% relative from ~30\% to close to 20\%. The maximum amount of data used for any one language was 200 hours of speech (Turkish).  Larger gains could probably be observed by fine tuning with significantly more training data, computing and people resources. 

Is  this level of performance (~20\% WER) adequate for accurate extraction of cultural markers? It is difficult to answer this question directly. Unlike speech recognition tools such as Whisper and WhisperX, there are no controlled data bases and open source code for extraction of spoken cultural markers that would permit straightforward evaluations  as a function of speech recognition error rates. 

One (admittedly, crude) proxy for a full evaluation is the evaluation of typical NLP components for spoken language processing as a function of Word Error Rate. Such studies are rare (and multilingual studies basically nonexistent). In one such recent study, synthetic speech corrupted by varying levels of noise was used to evaluate the impact of WER degradation on 3 different NLP components \cite{shapira-etal-2025-measuring}.  

The components examined were: Summarization, Question Answering (Q\& A) and Dialog Act Classification (DAC). Several measures of sensitivity to transcription quality were examined, but most germane to this discussion is the metric they called the {\it noise toleration point} (NTP). The NTP is the increase in WER that generates a "statistically significant" degradation in task metric relative to baseline no-noise performance. Here "statistical significance" is defined as a performance degradation by one standard deviation as estimated from metric score statistics. Each component was implemented with four different models corresponding to four different LLMs producing four different NTP values per component.  

\begin{table}[tb]
    \centering
    \caption{Word Error Rate Degradation Tolerances Estimated for Three NLP Components (from \cite{shapira-etal-2025-measuring})}
    \centering
    \label{tab:ntp}
    \begin{tabular}{lr}
        \toprule
        \textbf{Metric} & \textbf{WER Tolerance (NTP)} \\
        \midrule
        Summarization  & 7\%-30\% \\
        Q\&A & 5\%-34\% \\
        Dialog Act Classification & 44\%-71\% \\
        \bottomrule  \\
\end{tabular}
\end{table}

Table \ref{tab:ntp} contains the range of NTP values for each component. It can be seen that DAC is the least sensitive to increases in WER, with a range of 44\%-71\% across models. Insofar as DAC is analogous to topic spotting, this does not come as a complete surprise; it has been known since the early 1990s \cite{10.3115/1075671.1075697} that good topic spotting could be done at high word error rate levels. Summarization and Q\&A seem more sensitive to model quality;  some models start degrading even for very low word error rates while others seem more robust, at least up WER ranges of 20\%-30\%. 

It would therefore appear that out of the box Whisper or WhisperX performance for the languages investigated here (with  word error rates $\sim$30\%) might somewhat degrade NLP component performance, but that fine tuning with as little as 50-100 hours of data might lower WERs to 20\%- the point of producing results close to what would be achieved with perfect transcriptions. Of course, this is highly speculative given that underlying NLP component performance may vary across languages and that the effects of degradations on multiple NLP components may have a multiplicative effect on performance of cultural marker performance extraction systems built upon such components. 

It should be noted that true word error rates are not being measured, just deviations from closed captions of indeterminate quality that were provided without documentation. However, the WERs do decrease substantially after both  fine-tuning and inference parameter tuning. This suggests that the transcripts are not completely inaccurate. Nevertheless, it certainly would have been desirable to have gold-standard transcripts generated for the test data for these datasets. 

\section{Data Release}
\label{release}
One of the goals of this work was to create a set of processes, data and tools that would make it easier for non-speech-experts to generate speech transcripts for the purpose of identifying cultural markers. The original plan was to release all the data described in this paper, as the data selection process using the YouTube API (Section \ref{sec:pipeline}) specified to filter on the basis of the presence of a Creative Commons license. Unfortunately, in the data release process, it was found that although the filter in the YouTube API was set to select only Creative Commons data, there was no actual license information in the video metadata for about half the data, and some fraction of the videos were no longer available (Table \ref{tab:cc}. 

\begin{table}[th]
    \centering
    \caption{Creative Commons License Present in Video Metadata }
    \centering
    \label{tab:cc}
    \begin{tabular}{lrrrr}
        \toprule
        \textbf{Language} & \textbf{Original} & \textbf{Present} & \textbf{Absent} & \textbf{Video Vanished} \\
        \midrule
        Mandarin & 158 & 130 & 6 & 22 \\
        Spanish & 33 & 0 & 33 & 0 \\
        Russian & 204 & 1 & 199 & 4 \\
        Japanese & 101 & 1 & 100 & 0 \\
        Korean & 283 & 211 & 1 & 71 \\
        Turkish & 827 & 441 & 14 & 372 \\
        Hebrew & 156 & 148 & 0 & 8 \\
        \bottomrule  \\
\end{tabular}
\end{table}

Therefore, to err on the side of caution for now, only a subset of the data can be released. Three of the languages (Spanish, Russian, and Japanese) had little or no license information explicitly set in the video metadata. The release \cite{michael_picheny_2026} 
contains both the speech data, and a set of of tsv files breaking the data into train, test, and dev subsets. The speech data is provided after the segmentation described in Section \ref{sec:pipeline} was performed, along with the transcriptions of all the speech segments. 

\begin{table}[th]
\centering
\caption{Statistics on the Released Data}
\centering
\label{tab:multidir_lang_runs_summary_flipped}
\begin{tabular}{lrrrr}
\toprule
\textbf{Metric} & \textbf{Korean} & \textbf{Mandarin} & \textbf{Hebrew} & \textbf{Turkish} \\
\midrule
Train Segments & 41865 & 25930 & 59128 & 102615 \\
Train Hours & 43.48 & 30.42 & 61.38 & 105.07 \\
Dev Segments & 1113 & 1402 & 5255 & 7285 \\
Dev Hours & 1.09 & 2.88 & 5.16 & 7.74 \\
Test Segments & 3027 & 5250 & 6558 & 13847 \\
Test Hours & 3.44 & 6.42 & 7.28 & 13.61 \\
Error \% (No-Train) & 32.1 & 21.4 & 27.8 & 29.4 \\
Error \% (Train) & 23.4 & 10.2 & 22.1 & 18.2 \\
\bottomrule
\end{tabular}
\end{table}

Table \ref{tab:multidir_lang_runs_summary_flipped} summarizes the statistics on the released data. Error rates are presented before and after fine-tuning using the  Whisper large-v2 model with the optimized configurations described in Section \ref{sec:experiments}. Even though the amounts of training data are less than the amounts in the original configurations, significant improvements after fine-tuning are still observed for all languages. 

\section{Summary}
\label{sec:summary}
 A methodology for researchers interested in cross-cultural understanding to generate new data for input into training systems for cross-cultural understanding and other natural language tasks was described. The methodology proposed does not require deep speech recognition expertise and leverages open source data and tools. It was shown that even with relatively small amounts of fine tuning data (30-200 hours), one could obtain substantial improvements in speech recognition performance using state of the art transcription tools such as Whisper and WhisperX when manual transcriptions are available. In addition, 260 hours of training, dev, and test data were released to the community in Mandarin, Korean, Turkish, and Hebrew that can be used for future research by the community to improve speech recognition performance for cultural understanding research. Future work in this area might include experimenting with more sophisticated fine tuning methodologies, or perhaps ensembling multiple multilingual speech recognizers to achieve further improvements in performance.  In addition, investigating the actual impact of WER on cross-cultural information extraction would be fascinating when such automated tools become available to the broader community. 

\section*{Acknowledgment}
The author would like to thank Profs. He He and Kyunghyun Cho of NYU for their advice and support and the following NYU students who were part of the CCU program. They created and ran the processing pipeline that extracted speech transcriptions for the many evaluations in the actual program: Nikhil Verma, Sukrit Rao, Jash Rathod, and Sriphani Bellamkonda. 

AI assistance (ChatGPT 4.3) was used to help create properly formatted Tables \ref{tab:langstats} and \ref{tab:multidir_lang_runs_summary_flipped} and Figures \ref{fig:oob} and \ref{fig:ft}. In addition, it was consulted for for suggestions on how to reduce memory footprint and improve i/o performance for the original training code. Claude Sonnet 4.6 suggested the use of bootstrap methods for the significance tests, produced code and ran the tests described in the Results section. 

\bibliographystyle{IEEEtran}
\bibliography{slt2016}

\end{document}